%% file: prl.tex
\documentclass[aps,prl,twocolumn,showpacs,superscriptaddress,groupedaddress]{revtex4}  % for review and submission
\pdfoutput=1
\usepackage{graphicx}  % needed for figures
\usepackage{dcolumn}   % needed for some tables
\usepackage{bm}        % for math
\usepackage{amssymb}   % for math
\begin{document}

% The following information is for internal review, please remove them for submission
\widetext
\leftline{Version 1.0 as of \today}
\leftline{Primary authors: W. Chen}
\leftline{To be submitted to Physical Review Letters }
%\leftline{Comment to {\tt d0-run2eb-nnn@fnal.gov} by xxx, yyy}
%\centerline{\em D\O\ INTERNAL DOCUMENT -- NOT FOR PUBLIC DISTRIBUTION}

% the following line is for submission, including submission to the arXiv!!
%\hspace{5.2in} \mbox{Fermilab-Pub-04/xxx-E}

\title{EGAM Induced by Energetic-electrons and Nonlinear Interactions among EGAM, BAEs and Tearing Modes in a Toroidal Plasma}

\input author_list.tex       % D0 authors (remove the first 3 lines
                             % of this file prior to submission, they
                             % contain a time stamp for the authorlist)
                             % (includes institutions and visitors)
\date{\today}

\begin{abstract}
In this letter, it is reported that the first experimental results are associated with the GAM induced by energetic electrons (eEGAM) in HL-2A Ohmic plasma. The energetic-electrons are generated by parallel electric fields during magnetic reconnection associated with tearing mode (TM). The eEGAM localizes in the core plasma, i.e. in the vicinity of q=2 surface, and is very different from one excited by the drift-wave turbulence in the edge plasma. The analysis indicated that the eEGAM is provided with the magnetic components, whose intensities depend on the poloidal angles, and its mode numbers are $\mid$m/n$\mid$=2/0. Further, there exist intense nonlinear interactions among eEGAM, BAEs and strong tearing modes (TMs). These new findings shed light on the underlying physics mechanism for the excitation of the low frequency (LF) Alfv{\'e}nic and acoustic fluctuations. 
\end{abstract}

\maketitle

%\section{\label{sec:level1}First-level heading}
% sections are not used for PRL papers

\textbf{\textit{Introduction}}--The very low-frequency (LF) Alfv{\'e}nic and acoustic fluctuations, such as beta-induced Alfv{\'e}n eigenmode (BAE), and geodesic acoustic mode (GAM), are presently of considerable interest in the present-day fusion and future burning plasmas \cite{aff}, e.g. ITER. The low-frequency waves can significantly affect the plasma performance, and induce the particle losses and reduce the plasma self-heating. These LF instabilities can play an important role in turbulence and anomalous transport regulation, especially, while there is significant fraction of high energy particles in plasma \cite{phd}\cite{afu}. They can be used as energy channels to transfer the fusion-born-alpha-particle energy to the thermonuclear plasma, i.e. GAM/BAE channeling \cite{msa}.

The GAM with toroidal mode number n=0 is an eigenmode sustained by the coupling of radial electrostatic field and the poloidal variational density perturbations, and is usually taken to be electrostatic oscillation. The GAM is excited via modulation instability and pumped by the nonlinear interaction of drift wave turbulence \cite{phd}, and also driven by fast ions \cite{fuprl}\cite{naprl}\cite{qiuzy}. The GAM was investigated both using gyro-kinetic simulations and analytical methods in toroidal and slab geometry, and observed extensively in torus plasma \cite{phd}\cite{afu}. Meanwhile, the BAE with $n\neq 0$ is also a low frequency mode with parallel wave number $k_\|=(n-m/q)/R_0=0$ , which is due to the plasma finite beta effect under the geodesic curvature, and usually believed to be electromagnetic oscillation, and created by the coupling between the shear Alfv{\'e}n continuum with the poloidal mode number m and the sound continuum with the mode numbers m-1 and m+1, and driven by fast particles or large magnetic island. The BAEs were observed and investigated under different conditions in tokamak plasma \cite{chenw11}.

It is worthwhile noting that the BAE and GAM have similar dispersion relations in the case of the long wavelength limit, i.e., the kinetic expression of the GAM dispersion relation can degenerate with that of the LF shear Alfv{\'e}n accumulation point (BAE) \cite{fzppcf98} \cite{fzcl06}, which is useful for helping reciprocally identify the instabilities in the experiments. The most simple dispersion relations of BAE/GAM are given by
\begin{displaymath}
\omega_{BAE}=\omega_{GAM}\approx(2T_i/m_i)^{1/2}(7/4+T_e/T_i)^{1/2}/qR_0
\end{displaymath} 
Where q is safety factor, $R_0$ is major radius, $m_i$ is ion mass, and $T_i$, $T_e$ are ion and electron temperatures, respectively.
The energetic-electrons and magnetic-island induced BAEs had been observed and investigated on HL-2A in the previous works \cite{chenwprl} \cite{chenw11}. In this letter, it is reported that the first experimental results are associated with the GAM induced by energetic-electrons (eEGAM), and also present that there exists the intense nonlinear interactions among eEGAM, BAEs and strong TMs. 

\textbf{\textit{Experimental conditions and mode characteristics}}--HL-2A is a medium-size tokamak with major/minor radius $R/a=1.65m/0.4m$. The experiments discussed here were performed in deuterium plasmas with plasma current $I_p \simeq 150-170kA$, toroidal field $B_t \simeq 1.32-1.38T$, and safety factor $q_a \simeq 4.2-4.6$ at the plasma edge. The line averaged density was detected by a hydrogen cyanide interferometer. The polodial number m is measured using a set of seven Mirnov probes localized in the high field side (HFS) and eleven ones localized in the low field side (LFS). But the toroidal number n is measured using a set of ten Mirnov probes localized in the LFS of the vessel \cite{chenw10}. Four CdTe scintillator detectors are placed outside the vacuum vessel in order to obtain information of the hard x-ray emission, and chordal distances of sight lines  are $r_d$=5, 9, 15 and 30 cm, respectively. The range of the hard x-ray spectrum is $E_{\gamma}=10-200 keV$ divided into many energy bins by the PHA-software setting. 

\begin{figure}[!htbp]
\centering
\includegraphics[scale=0.7]{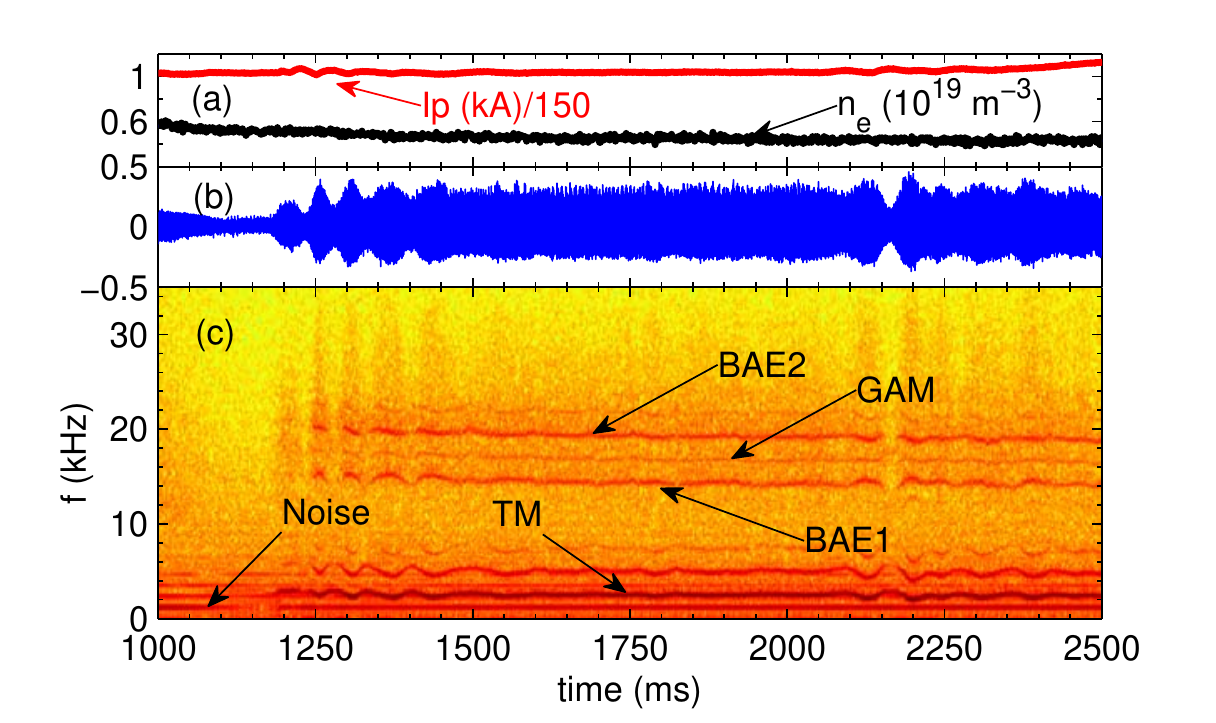}
\caption{\label{fig1} Experimental parameters of the typical discharge with strong TM on HL-2A. Plasma current, $I_p$ and density, $n_e$ (a), magnetic probe signal (b), and corresponding spectrogram (c), respectively.}
\end{figure}

The eEGAM has been observed in the HL-2A Ohmic plasma for the first time, recently. This phenomenon is perfectly reproducible, and a typical discharge parameters are shown in Fig.\ref{fig1}. A coherent MHD fluctuation is visible around 17.5 kHz from 1250 ms to 2500 ms. The toroidal mode number analysis indicates that this fluctuation does correspond to GAM due to n=0.  
In general terms, the magnetic component of GAM is two-order than the electric one, therefore it is very difficult that it is observed in Ohmic plasma. However, the magnetic components of GAM had been observed in the same discharge. The analysis indicated that the poloidal number of GAM is m=2, and the fluctuation intensity depends on the poloidal angles. The phenomena can be interpreted by Zhou's theory \cite{zhoud} which suggests that the GAM has a magnetic component with m=2, which is created by the m=2 parallel return current, and the fluctuation intensity depends on the poloidal angles, i.e., $\tilde{B}_{\theta} \propto sin(2 \theta)$. The similar experimental results ($\tilde{B}_{\theta} \propto sin(\theta)$), which are associated with the density fluctuation induced by GAM, can be found in the previous document \cite{akfprl}. 
The BAEs are also visible during strong TM activity with m/n=-2/-1 in the same discharge. The characteristics of the BAEs were investigated in previous works \cite{chenw11}. The mode numbers of the BAEs are m/n=2/1 and -2/-1. There exists an island width threshold ($\sim 3.4 cm$) for the BAE excitation on HL-2A \cite{chenw11}. Note that the BAEs can not be completely explained by the present theory \cite{abiprl}. The magnetic fluctuation spectrogram indicates that the GAM is always accompanied by strong TM and BAEs, and their frequencies comply with $f_{GAM}=f_{BAE2}-f_{TM}$, $f_{GAM}=f_{BAE1}+f_{TM}$ as well as $f_{GAM}=(f_{BAE2}+f_{BAE1})/2$. The GAM localizes in the core plasma, i.e., in the vicinity of q=2 surface where the ion Landau damping $\gamma_i$ is larger than the edge due to $\gamma_i \propto exp(-q^2)$, and it is very different from one excited by the drift-wave turbulence in the edge plasma on HL-2A \cite{zhaokj}\cite{lant}. Such GAM is not observed in the absence of strong TM or BAEs.

\begin{figure}[!htbp]
\centering
\includegraphics[scale=0.60]{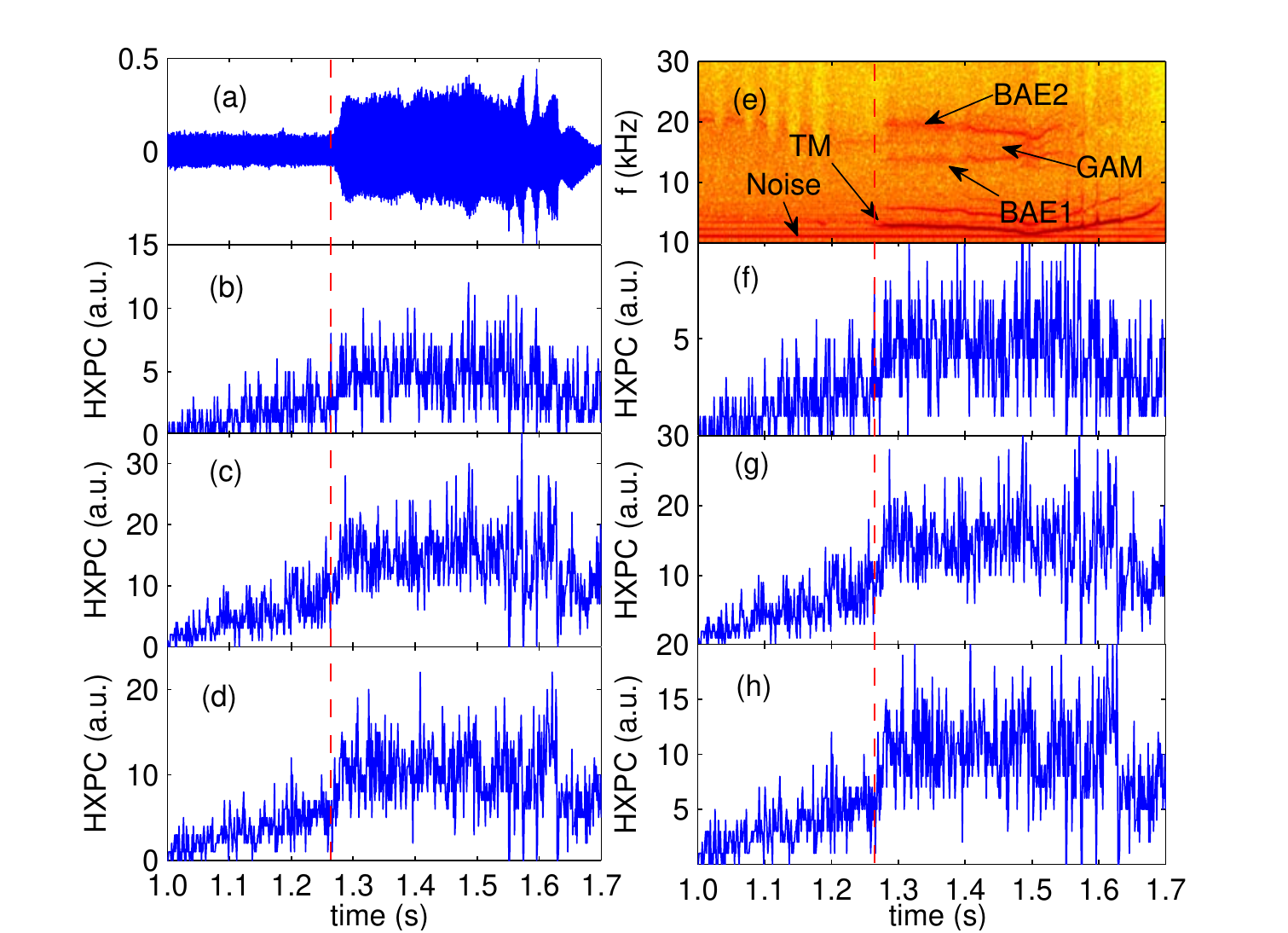}
\caption{\label{fig2} Enhancement of energetic electrons during magnetic reconnection at different CdTe channels on HL-2A for shot $\#$17455. Magnetic probe signal (a) and corresponding spectrogram (e). Hard X-ray counts in arbitrary unit, (b)-(d) and (f)-(h). Left column, $r_d=5 cm$; Right column, $r_d=30 cm$. (b) and (f), $E_{\gamma}=30-40keV$; (c) and (g); $E_{\gamma}=40-50keV$; (d) and (h), $E_{\gamma}=50-60keV$. Other energy bins do not been shown here.}
\end{figure}

\begin{figure}[!htbp]
\centering
\includegraphics[scale=0.63]{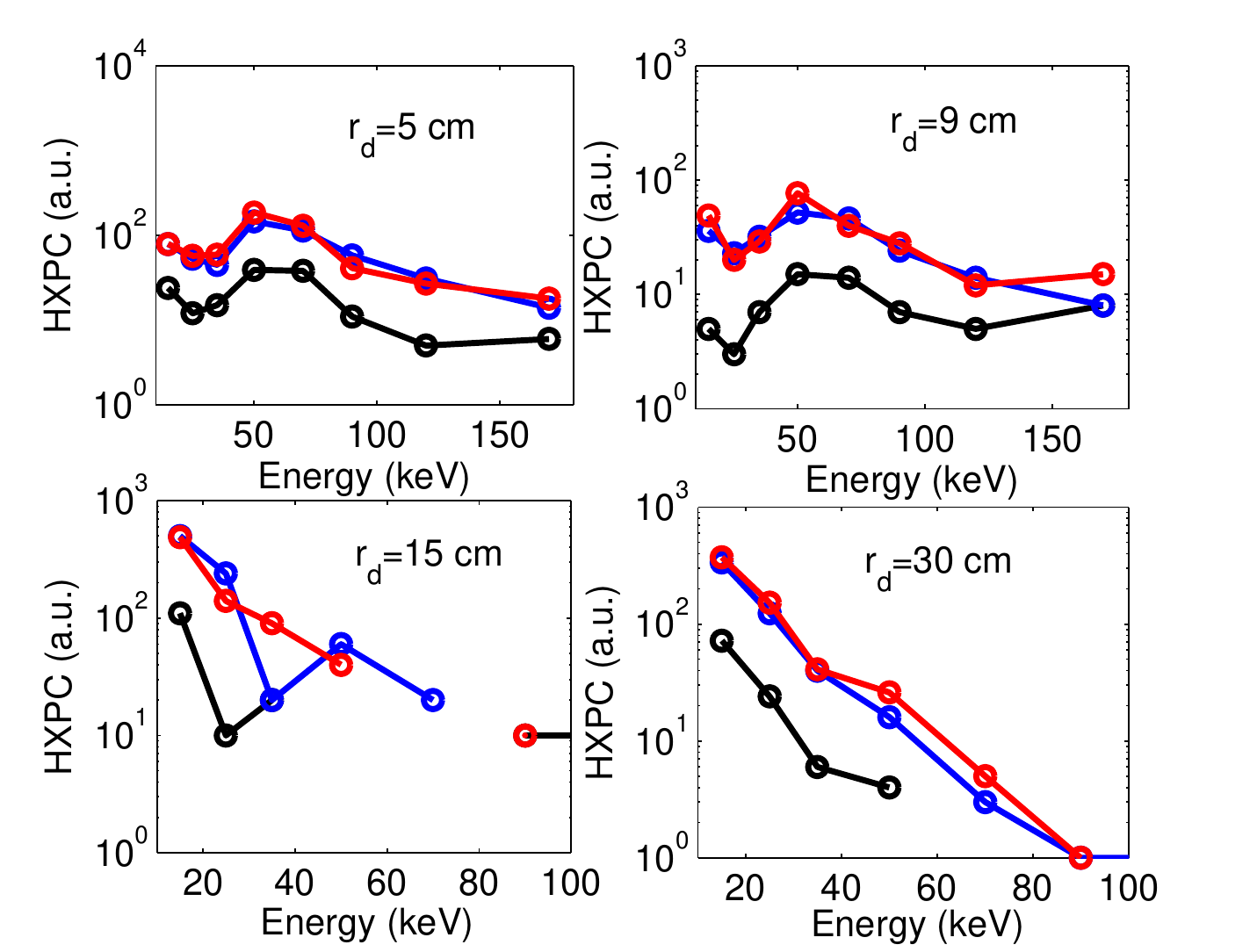}
\caption{\label{fig3} Energy distributions of energetic electrons without and with BAEs and eEGAM at different CdTe channels on HL-2A for shot $\#$17455. Black, blue and red lines are corresponding to t=1100-1110ms, 1300-1310ms and 1480-1490ms, respectively.}
\end{figure}

\textbf{\textit{Relationship between energetic-electrons and EGAM}}--The existence of energetic-electrons during magnetic reconnection results in the excitation of GAM. 
Generation of energetic-electrons during magnetic reconnection has been the subject of a number of theoretical and experimental investigations \cite{spvprl}\cite{djf}\cite{chenlj}\cite{ewa}. 
The production rate depend critically on the amplitude of the electric field generated during reconnection. The electric field is $E_{\|}=(s B_t / 16 r_s) w_m dw_m / dt$ \cite{spvprl}, where $w_m$ is the width of the magnetic island, $w_m = 4 (B_r r_s R_0 / n s B_t)^{1/2}$, and $dw_m / dt$ is the growth rate of magnetic island described by the tearing mode equation $dw_m / dt = 1.2 (\eta / \mu_0) \bigtriangleup^{\prime}_{m}$ in the case of low beta. Here, $\bigtriangleup^{\prime}_{m}$ is the stability parameter, $\eta$ is the plasma resistivity, $r_s$ is the radius of the magnetic surface, $B_r$ is the radial magnetic field perturbations, and $s=(r/q)dq/dr$ is the magnetic shear. On the basis of experimental parameters, we can evaluate that electric fields are of the order of $E_{\|} \sim 5V/m$ during the process of magnetic reconnection on HL-2A.
Analysis of HXR energy distribution has indicated that the energy of the energetic-electrons in flight is of the order of 20-200 keV. The time resolution of the PHA analysis did allow one to determine temporal modifications of the spectrum. More details will be introduced in a separate paper. Fig.\ref{fig2} shows that the HXR fluxes with different energy bins increase with TM growing at t=1270 ms, and the eEGAM is also driven. Further, during strong TM, the energy distributions of energetic-electrons are all enhanced at different CdTe channels, shown in Fig.\ref{fig3}, and the non-Maxwell distribution beams exist in the core plasma, as a result, these energetic-electrons induce the excitation of eEGAM.

\begin{figure}[!htbp]
\centering
\includegraphics[scale=0.55]{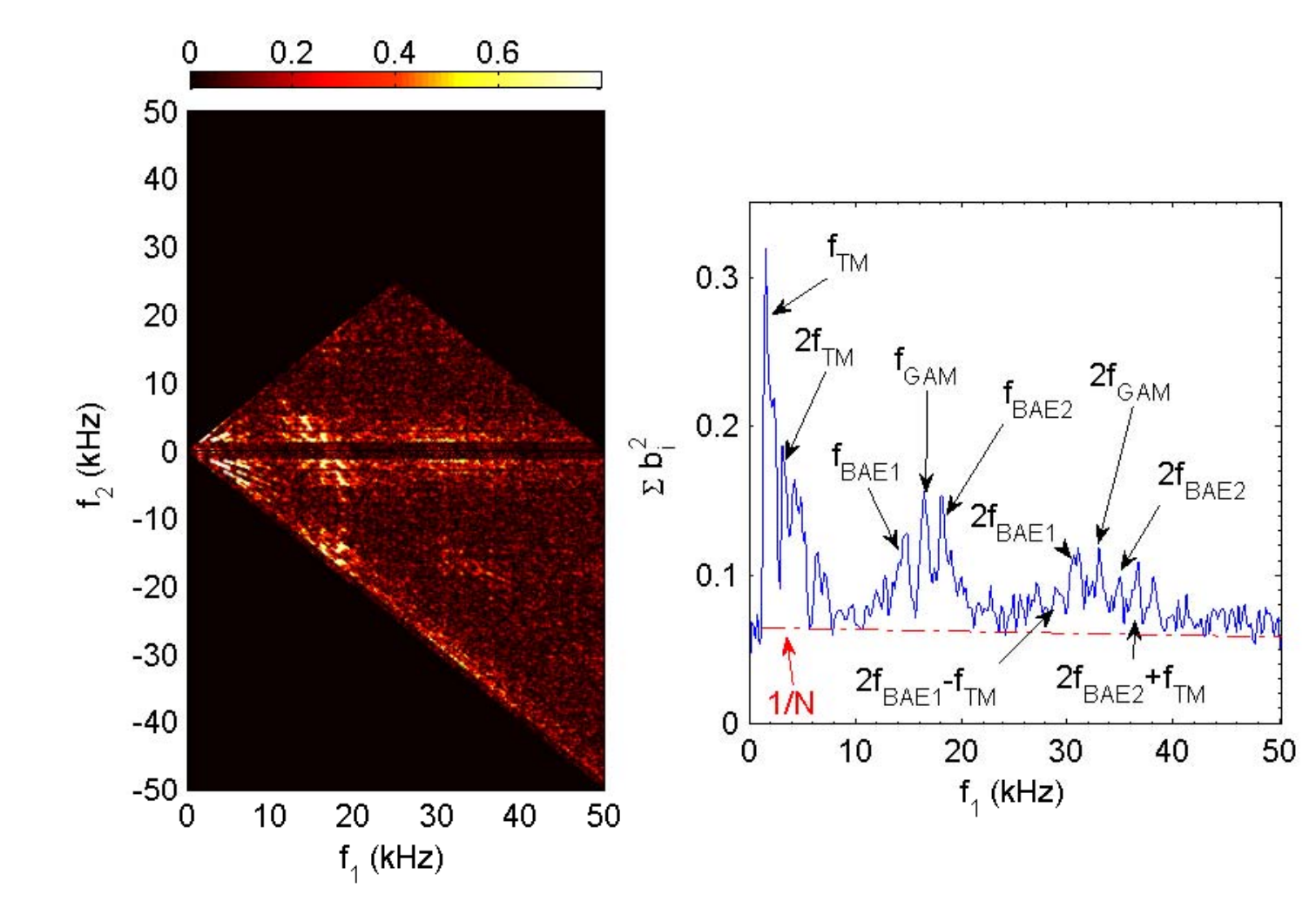}
\caption{\label{fig4} Squared bicoherence $\hat{b}^2_{\tilde{B}_{\theta} \tilde{B}_{\theta} \tilde{B}_{\theta}}(f_1,f_2)$ and summed squared bicoherence $\Sigma \hat{b}^2_{\tilde{B}_{\theta} \tilde{B}_{\theta} \tilde{B}_{\theta}}(f)$ of Mirnov signal at t=1.4-1.5 s for shot $\#$17455. Note that the resampling frequency is f=100 kHz, and nfft=512. 1/N denotes the noise level.}
\end{figure}

\textbf{\textit{Nonlinear interactions among BAEs, eEGAM and TMs}}--The nonlinear mode coupling can produce coherent mode structures which can provide overlap of wave-particle resonances in the minor radius, and transfer wave energy across different spatial scale. The role of nonlinear mode coupling is generally important in determining the mode excitation, saturation or damping. The nonlinear interaction also affects energetic particle redistribution/transport or plasma confinement. 
A novel result, which is nonlinear mode couplings among TM, BAEs and eEGAM, has been observed on HL-2A. For studying the nonlinear mode coupling, the squared bicoherence \cite{kyc} is given by $\hat{b}^2(f_1,f_2)=\mid \hat{B}_{XYZ} (f_1,f_2)\mid / <{\mid X(f_1) Y(f_2) \mid}^2> <{\mid Z(f_3) \mid}^2>$ with the Fourier bispectral $\hat{B}_{XYZ} (f_1,f_2)=<X(f_1) Y(f_2) Z^{\ast}(f_3)> $, $f_3=f_1 \pm f_2$ and $0 < \hat{b}^2(f_1,f_2) <1$. Where $X(f)$, $y(f)$ and $Z(f)$ are the Fourier transform of the time traces of $x(t)$, $y(t)$ and $z(t)$, respectively. The symbol $<>$ denotes the ensemble average over many realizations. It is convenient to represent the contribution of the nonlinear coupling from multiple modes to one mode with the summed squared bicoherence, which is defined as $\Sigma b^2_{XYZ}= \Sigma_{f = f_1 \pm f_2} \hat{b}^2(f_1,f_2) / N(f)$.  Where $N(f)$ is the numbers of realizations. To investigate the nonlinear mode coupling more precisely, the summed squared bicoherence must be higher than the noise level, which has the value of $(1-b^2_{XYZ} / N)$. Fig.\ref{fig4} shows the squared bicoherence and summed squared bicoherence of a poloidal Mirnov signal. It is found that the nonlinear interaction between the fundamental $n_{BAE}=1$ (or $n_{BAE}=1$ ) BAE with $f_{BAE}$ and $n=1$ TM with $f_{TM}$ at each different moment. The following matching conditions are satisfied among these modes, i.e. $n_{TM}+n_{BAE}=n^{\prime}_{BAE/GAM}$ and $f_{TM}+f_{BAE}=f^{\prime}_{BAE/GAM}$ for TM and BAEs. Moreover, the $n \pm 1$ BAEs interact with TM further and create a multitude of BAEs/GAMs, and $f^{\prime}_{BAE/GAM}=(k-1) \ast f_{TM}+f_{BAE}$, $n^{\prime}_{BAE/GAM}=(k-1)\ast n_{TM}+n_{BAE}$, where $k$ is positive integer. If only considering the fundamental frequencies of TM, BAEs and GAM, their nonlinear interaction relations are $f_{BAE2}-f_{BAE1}=2f_{TM}$, $f_{BAE2}+f_{BAE1}=2f_{GAM}$, and other expressions are $f_{BAE}=f_{GAM} \pm f_{TM}$ (i.e., $GAM+TM \Rightarrow BAE$) or $f_{GAM}=f_{BAE} \pm f_{TM}$ (i.e., $BAE+TM \Rightarrow GAM$). Need to point out here that the direction of wave energy transfer is unknown, and it needs to be assessed. The auto-bicoherence of magnetic fluctuation is shown that $\mid f_2 \mid = k \ast f_{TM}$ and $\mid f_2 \pm f_1 \mid = k \ast f_{TM}$ are strong coupling lines. It suggests that the nonlinear mode coupling process between TM and BAEs is similar with the coupling of drift-wave turbulence and GAM \cite{phd}\cite{afu} as well as TAE and EPM \cite{nac}. 

\textbf{\textit{Excitation Mechanism of eEGAM}}--The excitation mechanisms of eEGAM can be discussed, briefly. As we known, the energy transfer from energetic particle to the wave can be expressed by $G \varpropto \omega \partial f / \partial W + n \partial f / \partial P_{\varphi} $ Here, $\partial f / \partial W$ and $\partial f / \partial P_{\varphi}$ are, respectively,
the particle energy derivative of the distribution function of energetic particle and the derivative
for the toroidal momentum. For the GAM, the driving free energy may directly come from the positive
gradient of the distribution function $\partial f / \partial W$ and indirectly (because of n=0) from the radial derivative of distribution function by nonlinear interactions. The magnetic island induces BAEs, then the eEGAM is excited via the nonlinear interactions among BAEs and strong TM which is a pump wave. While the matching conditions are satisfied, i.e. $\omega_1+\omega_2 = \omega_3$ and $\vec{k}_1 + \vec{k}_2 = \vec{k}_3 $, the nonlinear interactions can occur between three waves in the plasma, and the coupled equations are bilinear and similar, and the coupling coefficients of the wave field amplitude determine the growth or damping of the waves \cite{stenflo}.  According to Chen's theory, while the purely Alfv{\'e}nic state described by the Wal{\'e}n relation ($\delta \vec{u}/ v_A = \pm \vec{B} / B_0 $) is broken \cite{chenl08}, i.e. $\delta E_{\parallel} \neq 0 $ or $\omega^2 \neq k_{\parallel}^2 v_A^2$, it will lead to significant perpendicular ponderomotive force and zonal flows. In our experiments, $\delta E_{\parallel} \neq 0 $ is satisfied obviously owing to the magnetic reconnection of strong TMs. The slow-sound-wave density and potential perturbation, which are induced by parallel ponderomotive, will have radially varying ($n = 0, m = \pm 1$) poloidal structures. The parallel ponderomotive force can couple with the compressible dynamics of slow-sound-waves, and the two high-frequency Alfv{\'e}nic modes can generate a low-frequency acoustic mode by nonlinearity. If the perpendicular incompressibility of shear Alfv{\'e}n wave is broken by the magnetic curvature, i.e. $\delta u_{\vert} \neq 0 $, the AEs can nonlinearly generate ($n = 0, m = \pm 1$) radially local magnetic perturbations. Recently, the theory investigation suggests that the plasma compressibility has a significant effect on nonlinear mode coupling of AEs, and the coupling of AEs is a more effective energy transfer at a lower amplitude level due to the enhanced compressional perturbations in the poloidal sidebands \cite{hirota}. The GAM is toroidally symmetric mode unique to toroidal plasmas with mode structure that is nearly poloidally symmetric. The spatial overlap of the mode structures should be essential for the nonlinear mode coupling. Our experimental results indicates the TM, BAEs and eEGAM all localize at the q=2 surface, i.e., there exists an overlap between the mode radial structures of eEGAM and BAEs. It means that the AEs can propagate poloidally into the region of the zonal flows (ZFs) due to the zonal mode structure of GAM and the mode structure overlap, and can interacts with the GAM/ZFs, and the wave energy can transfer between the GAM and AEs, and the intensity of the AEs can influence on the growth, saturation and damping of the GAM. 

\textbf{\textit{Summary}}--The eEGAM has been recently observed for the first time on HL-2A. The magnetic fluctuation spectrogram indicates that the eEGAM is always accompanied by strong TM and BAEs. The analysis reveals that the eEGAM is provided with the magnetic components, whose intensities depend on the poloidal angles, and its mode numbers are $\mid m/n \mid=2/0$. Further, a novel result, which is that there exist the cross-scale couplings among TM, BAEs and eEGAM, has been observed on HL-2A. The eEGAM is directly driven by energetic-electrons via the gradient of the velocity space or indirectly produced via the nonlinear mode coupling among BAEs and strong TM, but more theoretical works are needed because this phenomenon is a typical example with respect to multi-scale interactions. The eEGAM should have a significant effect on plasma transport in the vicinity of the magnetic island, and also have a profound regulatory effect on the turbulence around magnetic island. The experimental results indicate that the couplings possibly induce the energy transfer among TM, BAEs and eEGAM, and it is possible to be one of mechanisms of the energy cascade in Alfv{\'e}n turbulences, and the BAE/GAM may be an energy channeling between different scales, such as macro-, meso- and micro-scale. The new findings give a deep insight into the underlying physics mechanism for the excitation of the LF Alfv{\'e}nic/acoustic fluctuation and ZFs. 

The author (C.W.) is very grateful to the HL-2A Group, and thanks Prof. L. Chen, Dr. F. Zonca and Dr. G. Fu for their teaching me many theories in the past years, and also acknowledges Dr. Z. Qiu for valuable discussions. This work is supported in part by the National Natural Science Foundation of China under Grant No. 11005035 and 10935004.

\end{document}

%% file: author_list.tex
% remove these 3 lines before journal submittal.
%\centerline{author list dated 9 February 2012}
% end removal before journal submittal
%

\author{W. Chen$^1$, X.T.Ding$^1$, L.M. Yu$^1$, X.Q. Ji$^1$, J.Q. Dong$^1$, Q.W. Yang$^1$, Yi. Liu$^1$, L.W. Yan$^1$, Y. Zhou$^1$, W. Li$^1$, X.M. Song$^1$, S.Y. Chen$^2$, J. Cheng$^1$, Z.B. Shi$^1$, X.R. Duan$^1$ and HL-2A team}

\address{Southwestern Institute of Physics, P.O. Box 432 Chengdu 610041, China}
\address{College of Physical Science and Technology, Sichuan Univ., Chengdu 610065}

\noaffiliation
\vskip 0.25cm